\documentclass[preprint,journal]{vgtc}       % preprint (journal style)

\ifpdf%                                % if we use pdflatex
  \pdfoutput=1\relax                   % create PDFs from pdfLaTeX
  \usepackage{graphicx}                % allow us to embed graphics files
  \DeclareGraphicsExtensions{.pdf,.png,.jpg,.jpeg} % for pdflatex we expect .pdf, .png, or .jpg files
\else%                                 % else we use pure latex
  \usepackage{graphicx}                % allow us to embed graphics files
  \DeclareGraphicsExtensions{.eps}     % for pure latex we expect eps files
\fi%

\graphicspath{{figures/}{pictures/}{images/}{./}} % where to search for the images

\usepackage{microtype}                 % use micro-typography (slightly more compact, better to read)
\PassOptionsToPackage{warn}{textcomp}  % to address font issues with \textrightarrow
\usepackage{textcomp}                  % use better special symbols
\usepackage{mathptmx}                  % use matching math font
\usepackage{times}                     % we use Times as the main font
\usepackage{cite}                      % needed to automatically sort the references
\usepackage{tabu}                      % only used for the table example
\usepackage{booktabs}                  % only used for the table example
\usepackage{makecell}                  % only used for the table example
\usepackage{amsmath}
\usepackage{diagbox}
\usepackage{listings}
\usepackage[]{algorithm2e}
\usepackage{soul}
\usepackage{color}
\usepackage[dvipsnames]{xcolor}
\DeclareMathOperator*{\argmin}{argmin}
\DeclareMathOperator*{\sgn}{sgn}

\onlineid{1111}

\vgtccategory{Research}
\vgtcpapertype{algorithm/technique}

\title{Visual Analysis of High-Dimensional Event Sequence Data via Dynamic Hierarchical Aggregation}

\author{David Gotz, Jonathan Zhang, Wenyuan Wang, Joshua Shrestha, and David Borland}
\authorfooter{
\item
 David Gotz is with the School of Information and Library Science at the University of North Carolina at Chapel Hill. E-mail: gotz@unc.edu.
\item
 Jonathan Zhang is with the Dept. of Biostatistics at the University of North Carolina at Chapel Hill. E-mail: jzhang42@live.unc.edu.
\item
 Wenyuan Wang is with the School of Information and Library Science at the University of North Carolina at Chapel Hill. E-mail: vaapad@live.unc.edu.
\item
 Joshua Shrestha is with the Dept. of Computer Science at the University of North Carolina at Chapel Hill. E-mail: prayash@live.unc.edu.
\item
 David Borland is with RENCI at the University of North Carolina at Chapel Hill. E-mail: borland@renci.org.
}

\shortauthortitle{Biv \MakeLowercase{\textit{et al.}}: Global Illumination for Fun and Profit}
\abstract{
Temporal event data are collected across a broad range of domains, and a variety of visual analytics techniques have been developed to empower analysts working with this form of data.  These techniques generally display aggregate statistics computed over sets of event sequences that share common patterns.  Such techniques are often hindered, however, by the high-dimensionality of many real-world event sequence datasets which
can prevent effective aggregation. A common coping strategy for this challenge is to group event types together prior to visualization, as a pre-process, so that each group can be represented within an analysis as a single event type.  
However, computing these event groupings as a pre-process also places significant constraints on the analysis. This paper presents a new visual analytics approach for dynamic hierarchical dimension aggregation. The approach leverages a predefined hierarchy of dimensions to computationally quantify the informativeness, with respect to a measure of interest, of alternative levels of grouping within the hierarchy at runtime.  This information is then interactively visualized, enabling users to dynamically explore the hierarchy to select the most appropriate level of grouping to use at any individual step within an analysis.  Key contributions include an %efficient and tunable 
algorithm for interactively determining the most informative set of event groupings for a specific analysis context, % from within a large-scale hierarchy of event types, 
and a scented scatter-plus-focus visualization design with an optimization-based layout algorithm that supports interactive hierarchical exploration of alternative event type groupings.  
We apply these techniques to high-dimensional event sequence data from the medical domain and report findings from domain expert interviews.
} % end of abstract (400 words or less for IEEE VIS)


\CCScatlist{ % not used in journal version
 \CCScat{K.6.1}{Management of Computing and Information Systems}%
{Project and People Management}{Life Cycle};
 \CCScat{K.7.m}{The Computing Profession}{Miscellaneous}{Ethics}
}

\teaser{
  \centering
  \vspace{-0.1cm}
  \includegraphics[width=\linewidth]{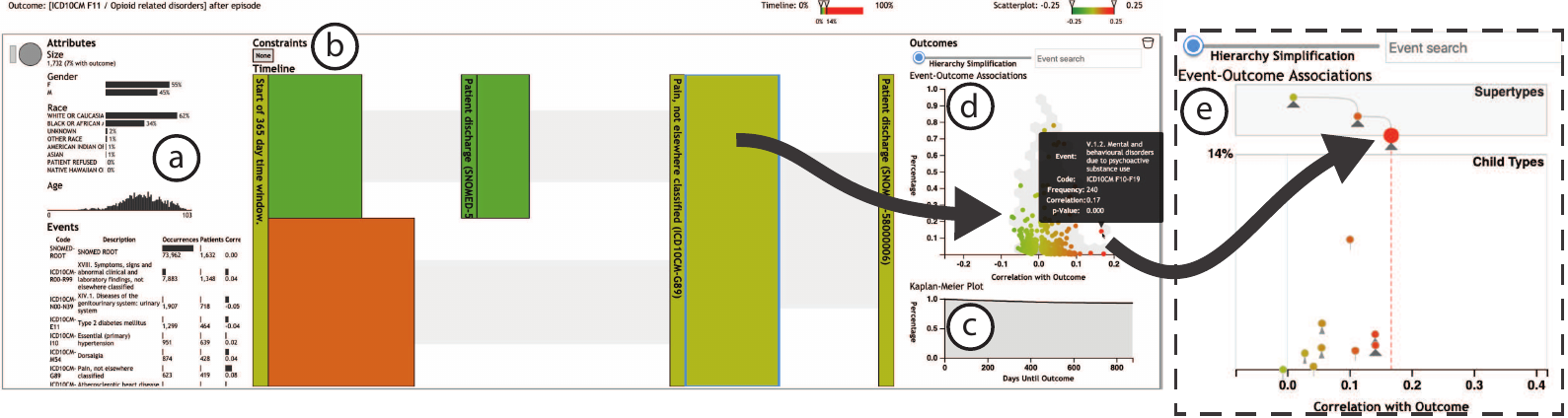}
  \vspace{-0.8cm}
  \caption{The Cadence system for temporal event sequence visualization. (a)~Interactive bar charts and histograms summarize non-temporal attributes and a variety of temporal event statistics. (b)~An interactive flow-based timeline allows users to dynamically define and explore pathways within the temporal event data. Selections within the timeline link to (c)~a Kaplan-Meier chart to summarize outcomes-over-time. Timeline selections also link to (d and e)~a novel scatter-and-focus visualization which leverages dynamic hierarchical aggregation, scenting, and optimization-based layout to support navigation of high-dimensional hierarchical event data.}
	\label{fig:teaser}
}

\vgtcinsertpkg

\begin{document}

\firstsection{Introduction}
\maketitle

\vspace{-0.1cm}
Temporal event data are collected and analyzed across a broad range of domains. Reflecting this ubiquity, visual analytics techniques have been designed to support event sequence analysis across a diverse set of application areas including sporting events (e.g., \cite{gotz_decisionflow:_2014}), career progression (e.g., \cite{guo_eventthread:_2018}), transportation logistics (e.g., \cite{guerra-gomez_analyzing_2011}), and large-scale system logs (e.g., \cite{wongsuphasawat_using_2014}). 
These applications typically aggregate and visualize large collections of event sequences--time-ordered lists of discrete events that describe some underlying process (e.g., an athletic team's performance over a season, an individual worker's career progression, the response by emergency services to a car accident, or a user's interaction with a website). By supporting the analysis of large numbers of event sequences describing the same underlying process, these tools enable users to gain insights about common patterns, rare event paths, and associations between temporal patterns and specific performance measures.

These analytical goals, however, are often hindered by the complexity present in real-world collections of event sequence data. For example, medical data analysis has been a widely studied application of event sequence visualization techniques (e.g., \cite{du_coping_2017,wongsuphasawat_outflow:_2011,wongsuphasawat_lifeflow:_2011,gotz_decisionflow:_2014,perer_frequence:_2014,perer_visual_2011,monroe_temporal_2013,rind_interactive_2013}).  Medical analyses routinely use data from large electronic medical record systems that contain data spanning several years for millions of patients \cite{raghupathi_big_2014,weber_finding_2014}.  Moreover, the data is represented using coding systems that have hundreds of thousands of unique types of events that can occur (diagnoses, medications, lab tests, etc.) (e.g., \cite{noauthor_icd_2018,snomed2019}).

Early visual analytics methods focused on overcoming challenges arising from a large volume of event sequences (as opposed to large numbers of distinct events types) by aggregating identical sequences of events, and visualizing aggregate statistics via tree-based or flow-based visualizations techniques.  This approach successfully enabled users to discover common patterns and their relative frequencies from a large number of sequences (e.g., \cite{wongsuphasawat_lifeflow:_2011,wongsuphasawat_outflow:_2011}).  However, these systems did not scale effectively to enable the analysis of high-dimensional event datasets with large numbers of distinct event types.  High-dimensionality hinders many aggregation-based approaches because the large number of event types produces an even larger number of distinct event sequence patterns.  Moreover, higher dimensionality typically results in increased sparsity, resulting in an increased number of smaller aggregated subgroups that limit the statistical significance associated with any patterns that are discovered. 

Responding to this challenge, more recent work has led to a variety of additional coping strategies to overcome the challenge of high-dimensionality. One widely used technique, which Du et al. found in 80\% (16 of 20) of the systems they surveyed \cite{du_coping_2017}, is \emph{grouping events by category}. In this strategy, systems define categories of events such that all occurrences of any of the distinct event types within a category are treated as equivalent.  This approach is highly effective because it directly addresses the dimensionality problem by limiting the number of event types.

However, this approach often requires very aggressive grouping (e.g., reducing hundreds of thousands of event types to dozens) and, as Du et al. observed \cite{du_coping_2017}, the grouping of events is typically performed as a pre-process (e.g., \cite{meyer_visualizing_2013,perer_data-driven_2013}).  This design choice--to decide as a pre-process which groups of event types should be treated as equivalent--is motivated in part by a very practical concern: most existing techniques have difficulty visualizing high-dimensional event sequence data.  Therefore, the number of event types must be reduced before data can be loaded into the visualization system.

Unfortunately, pre-defining which sets of events are treated as equivalent is highly constraining: (1) it prevents users from interactively using the visualization tools to look at event data from multiple levels-of-detail, and (2) it requires assumptions to be made at the time of pre-processing about which categorizations are most useful to reduce dimensionality for a given analysis task.  Moreover, 
the most appropriate level of grouping is both data- and task-dependent, suggesting that an analyst may be best served by different grouping choices at different points in the analysis process. 

Motivated by the need to provide users with interactive control over event type grouping as part of the event sequence analysis process, this paper presents a new visual analytics approach for dynamic hierarchical dimension aggregation. The approach leverages a pre-defined hierarchy of dimensions to computationally quantify the informativeness, with respect to a specific measure of interest (e.g., a medical outcome), of alternative groupings within the hierarchy.  This information is then interactively visualized, enabling users to dynamically explore the hierarchy to select at runtime the most appropriate level of grouping to use at any point during an analysis.  While these methods are not specific to medical data, we apply the techniques to high-dimensional event sequence analysis using large-scale event type hierarchies from the medical domain.

The key research contributions presented include:

\vspace{-0.3cm}
\begin{itemize}
\item An efficient and tunable algorithm that interactively determines, given user preferences, the most informative set of event groupings across a large-scale hierarchical set of event types for a given analytical context. The algorithm leverages a measure for quantifying the informativeness of a given event, or event grouping, with respect to a specific outcome measure of interest.
\vspace{-0.3cm}
\item A novel scatter-plus-focus visualization design that supports interactive hierarchical  exploration of the space of event type groupings. The visualization adopts scented navigation cues to help users navigate complex hierarchies.  Interactive focus control and an optimization-based layout algorithm are used to manage complexity and overcome challenges of overplotting.
\vspace{-0.3cm}
\item Integration of the above methods within \emph{Cadence}, a web-based visual analytics system for population health applications that enables users to explore high-dimensional temporal event sequence datasets using dynamic hierarchical aggregation.
\end{itemize}
\vspace{-0.3cm}

This paper describes the contributions enumerated above, presents an example use case, and reports findings from domain expert interviews. A discussion of these results highlights strengths of the dynamic hierarchical aggregation techniques described in this paper and motivates opportunities for future research.
\section{Related Work}

The contributions of this paper are most closely related to past research on event sequence visualization, event sequence analysis, hierarchical visualization, and visual scenting.  In addition, the prototype application used for the case study and interviews, \emph{Cadence}, leverages techniques for tracking provenance and selection bias.

\subsection{Visualization of Temporal Event Sequences}

Given the broad range of applications, as well as the unique methodological challenges posed by the volume and complexity of event sequence data, the visualization of temporal event sequences has been widely studied.  Initial approaches focused on interactive alignment around sentinel events with individually drawn event sequences (e.g., \cite{wang_aligning_2008}).  These techniques were powerful, but had limited utility when working with large numbers of event sequences, as is common in many real-world applications.

Later work utilized aggregation to visualize large volumes of event sequences.  This approach enabled the visualization of datasets with very large numbers of event sequences by computing aggregate data structures in tree \cite{wongsuphasawat_lifeflow:_2011} or graph \cite{wongsuphasawat_outflow:_2011} form, then rendering visual representations of these structures (e.g, using icicle plots \cite{kruskal_icicle_1983} or Sankey-like diagrams \cite{riehmann_interactive_2005}, respectively) rather than individual events. 

These aggregation-based approaches are, in theory, infinitely scalable for large numbers of individual sequences (in the same way that a bar chart can be used equally well to visualize a binary categorical distribution for 100 items or 1 billion items).  However, the variety of data contained within many event sequence datasets--which can contain upwards of tens of thousands of unique event types--is more fundamentally challenging for visualization because the resulting increase in variation between event sequences interferes with aggregation.

This challenge has led to a variety of approaches exploring alternative coping strategies \cite{du_coping_2017}, including a priori event selection (choosing a small number of events to include in an analysis and ignoring the rest, e.g., \cite{wongsuphasawat_using_2014}), dynamic event selection via user interaction (e.g., \cite{gotz_decisionflow:_2014}), 
simplification by pattern substitution (e.g., \cite{monroe_temporal_2013}), and event replacement rules that account for event attributes \cite{cappers_exploring_2018}.  
Additional strategies related to hierarchies are reviewed in Section \ref{sec:related_hierarchies}.  These approaches typically require users to select which events to include within an analysis without any grouping of similar events, or to manually identify patterns/rules to use for substitution based on an understanding of the non-simplified data.
Of these approaches, the methods in this paper are closest to dynamic event selection \cite{gotz_decisionflow:_2014}.  However, rather than simply selecting which events to include, we leverage hierarchical information to support dynamic aggregation of similar event types.  %

\subsection{Event Sequence Analytics}

The use of computational methods to help surface statistically interesting patterns or features within event sequence data has been widely explored.  Moreover, the results from these efforts show that combining computational approaches with interactive user exploration can be highly effective \cite{malik_high-volume_2016}. In some cases, pattern mining algorithms have been deployed to help identify sequential event patterns with high frequency \cite{perer_frequence:_2014} or with strong associations to some attribute of the sequences (e.g., an outcome measure) \cite{gotz_methodology_2014}.

Analytics have also been computed at runtime in a recursive fashion in response to user interaction.  This approach has been used to support high dimensional event sequence visualization through dynamic event selection in DecisionFlow \cite{gotz_decisionflow:_2014}, to enable interwoven queries and mining that can be performed at any point within a visual analysis in MAQUI \cite{law_maqui:_2018}, and to iteratively identify groups of similar sequences \cite{unger_understanding_2018}.

The methods and prototype described in this paper adopt a similar recursive analytics approach, including dynamic event selection as in DecisionFlow and enabling multiple panels of inquiry as in MAQUI.  However, the primary research contributions focus on a different challenge: recursive analytics to help users select effective event groupings (i.e., groups of event types to treat as equivalent).

Other computational approaches have focused on event sequence comparison, including projects that have compared 
event sequences for clickstreams (e.g., \cite{zhao_matrixwave:_2015}) and medical patient cohorts (e.g., \cite{malik_high-volume_2016,malik_cohort_2015}). 
The prototype system described in this paper %
exhibits some similarities to these projects, but they are not directly relevant to the research contributions outlined in this paper. 

\subsection{Hierarchies}
\label{sec:related_hierarchies}

Hierarchical aggregation is widely used in visualization \cite{elmqvist_hierarchical_2010}, including levels-of-detail within graph drawing (e.g., \cite{zinsmaier_interactive_2012}) and topic evolution in text visualization (e.g., \cite{cui_how_2014}).  
Not surprisingly, therefore, Du et. al's survey of event sequence visualization methods (which routinely employ graph- or tree-based visual representations) found that hierarchical aggregation is often used as a strategy for coping with complexity in event sequence visualization \cite{du_coping_2017}. However, as Du et. al also recognized, the process typically relies on a pre-process to determine how to aggregate data, without any input from the user of the visualization and without accounting for the context of a given analysis as it unfolds. 

For example, CareFlow grouped medication data into predefined classes of drugs prior to visualization \cite{perer_data-driven_2013}. Similarly, ScribeRadar mapped attributes of events to a hierarchy of event types for visualization via an icicle plot using a pre-processing step that maps raw events into pre-defined hierarchical categories \cite{wongsuphasawat_using_2014}.
In contrast, this paper describes an approach that enables interactive, user-guided aggregation at runtime using hierarchical type relationships.   

While the methods of this paper are broadly applicable, the %
described prototype system is designed to support event sequence analysis within the medical domain.  Reflecting this, we leverage existing medical coding hierarchies including ICD-10-CM \cite{noauthor_icd_2018} and SNOMED-CT \cite{spackman_snomed_1997}. These medical coding systems together include over 100,000 distinct types of medical events (e.g., diagnoses, procedures), providing a specificity that is often too fine-grained to support effective analysis without event type grouping.  This is reflected by major efforts within the health informatics community to better understand and manage event groupings during analysis \cite{jiang_building_2017,gold_clinical_2018,bodenreider_nlm_2013}.  

Visual navigation of hierarchies (e.g., \cite{landesberger_visual_2011}) is also closely related to the work presented in this paper.  This includes techniques for navigating large and complex hierarchies or graphs using focus+context methods (e.g., \cite{stasko_focus+context_2000, tominski_fisheye_2006}) and techniques that focus on local neighborhoods or paths (e.g., \cite{van_ham_search_2009,partl_pathfinder:_2016}).  In the spirit of these techniques, the scatter-plus-focus visualization proposed in this paper uses local hierarchical structures (the path to the hierarchy root, plus all children) to enable efficient interactive navigation of the event type hierarchy.

\subsection{Scenting, Provenance, and Selection Bias}

A number of other related areas of research have also informed the work presented in this paper including scenting, provenance, and selection bias. Scenting is an interface technique that provides users with visual cues that help direct their interactions toward more informative subspaces of information \cite{olston_scenttrails:_2003}.  Scented Widgets follow this approach by adorning traditional widgets (e.g., radio buttons and sliders) with visualization-based cues to facilitate informed navigation within information spaces \cite{willett_scented_2007}.  The scatter-plus-focus visualization proposed in this paper similarly adopts a scent-based approach to help users navigate large event grouping hierarchies.

The \emph{Cadence} prototype, meanwhile, includes features that track user-defined cohorts over time both (1) as a record of insight provenance and (2) as a means to track and communicate information about emerging selection bias during the cohort selection process.  These features build upon prior work in both visual cohort selection \cite{zhang_iterative_2015} and selection bias \cite{gotz_adaptive_2016,gotz_adaptive_2017}.  However, while these features are part of %
the system, they are beyond the scope of this paper and described elsewhere \cite{borland_selection_2020}.

\section{Requirements for Hierarchical Aggregation}

High-dimensional temporal event sequence data can be challenging to analyze for multiple reasons.  First, the large number of distinct event types produces a high amount of variance between sequences.  This inhibits effective aggregation of similar sequences, a key step in the visualization process for scalable event sequence visualization techniques.  Second, the sparse and fine-grained specificity of high-dimensional event data can mask patterns of interest by introducing variation between sequences due to distinct event types that should in fact be treated as equivalent for a given analysis task. 

Fortunately, high-dimensional event data is rarely composed of entirely independent event types. Instead, metadata is often available (or can be defined) to organize events within a hierarchy that captures relationships between event types (e.g. see Section~\ref{sec:related_hierarchies}).
The section presents examples from real analysis tasks in the medical domain to highlight both the challenges of high-dimensional event data and the benefits of hierarchical metadata.  This section then concludes with a series of design requirements that motivate the methods in Section~\ref{sec:methods}.

\subsection{Medical Events}

As outlined in the introduction, medical analyses are a common focus of temporal event sequence visualization tools.  This is due in part to the fact that medical institutions capture large amounts of event data about patients over time as part of the normal care delivery process.  
Much of this data is entered into electronic health record systems (EHRs) which leverage a variety of standardized coding systems such as ICD-10-CM \cite{noauthor_icd_2018} and SNOMED-CT \cite{snomed2019}.  These coding systems provide common representations that enable the interchange of data between information systems (e.g., doctor-facing EHRs and patient portals) or parties (e.g., for medical billing), and are often used for retrospective analysis \cite{rea_building_2012,jensen_mining_2012,hersh_recommendations_2013}.  

Medical coding systems include \emph{hundreds of thousands} of unique types of events (e.g., diagnoses, procedures, medications).  Moreover, given the large number of distinct event types, the data can be very sparse with many events occurring rarely or not at all even within large datasets.  In addition, there can be significant variation in the way medical events are coded using distinct but semantically similar events.

\begin{figure}[t]
  \centering
  \includegraphics[width=2.1in]{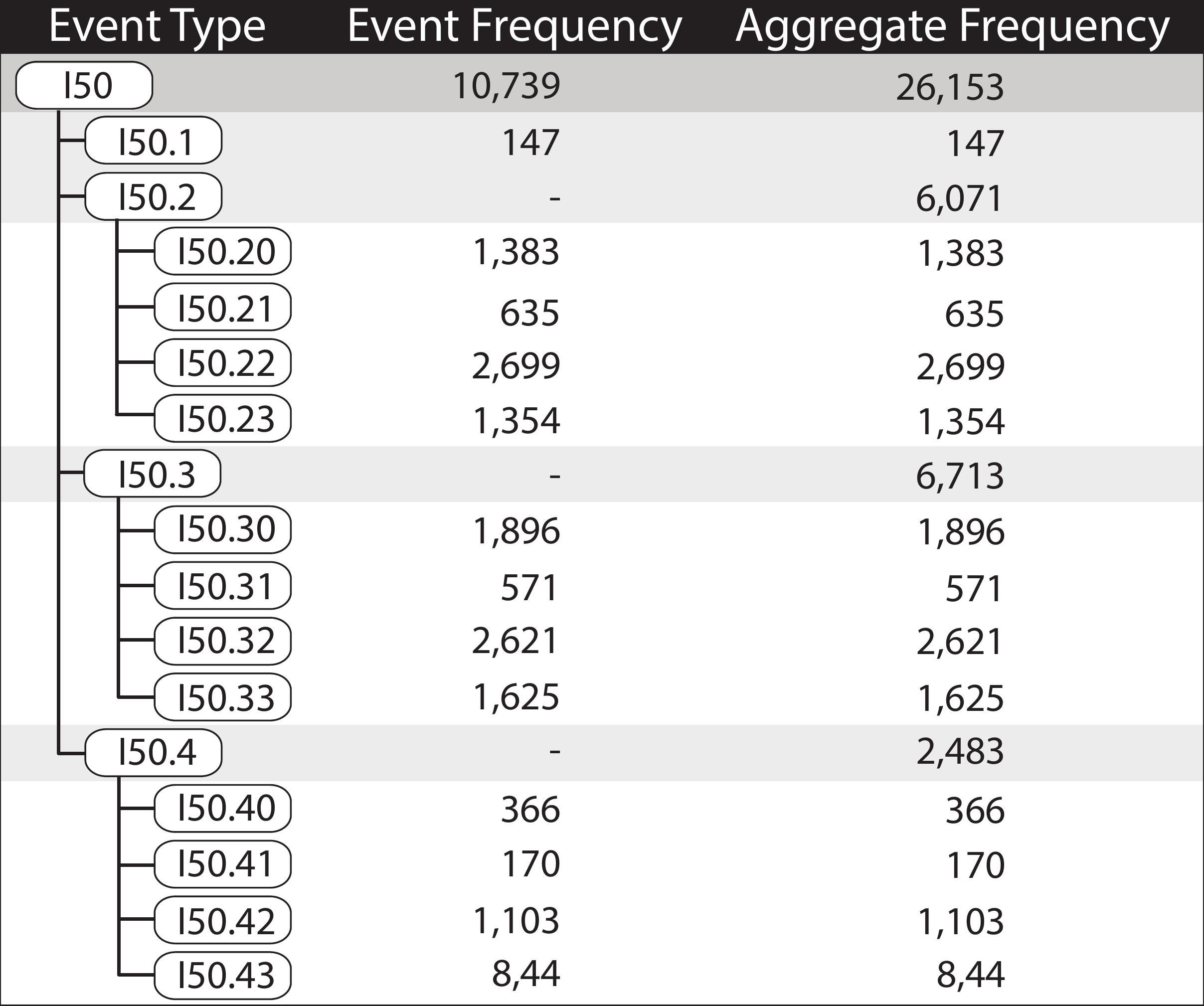}
    \vspace{-0.35cm}
    \caption{In a dataset of 16,983 diabetes patients, 5,084 also had a heart failure diagnosis (ICD-10 codes with the I50 prefix).  A total of 26,153 heart failure codes events were observed
    across 14 different variants. %
    }
\label{fig:i50data}
    \vspace{-0.35cm}
\end{figure}

For example, consider the statistics reported in Figure~\ref{fig:i50data}, which summarize the occurrence frequencies of various forms of Heart Failure (the I50 family of ICD-10 codes) within a dataset from UNC Health Care containing 16,983 diabetes patients. Nearly a third of that cohort--5,084 patients--were also diagnosed with heart failure.  As shown in the figure, heart failure was represented in the dataset by one of several distinct codes (I50, I50.1, I50.20, etc.)  These codes capture various forms of heart failure, e.g., diastolic, acute on chronic systolic, and several others.  In all, the dataset contains 26,153 occurrences of heart failure events, with about 40\% (10,739) represented as generic heart failure (ICD-10 code I.50).  The remaining events are spread out across 13 other codes. 

As Figure~\ref{fig:i50data} shows, however, there is a hierarchical relationship between the ICD-10 codes.  Depending on the circumstances, an analyst might wish to treat all versions of heart failure (all variants of the I50 code) as equivalent.  At another point in the same analysis, the user may wish to distinguish between different subgroups of heart failure (I50.1 vs. I50.2 vs. I50.3 vs. I50.4) to enable comparison between subgroups.  After finding something interesting in one of these groups (e.g., I50.4), an analyst may wish to compare the other I50.1/I50.2/I50.3 codes against a more detailed breakdown of I50.4 (i.e., I50.40, I50.41, I50.42, and I50.43).  

This example demonstrates how single conceptual values (e.g., Heart Failure) may be spread out across multiple event types that can be aggregated in many ways depending on the semantics of a particular analysis.  Critically, however, this example discusses just 17 event types.  As shown in Table \ref{tab:querystats}, cohorts returned for realistic queries against real-world medical data can contain well over 10,000 unique event types.  With an event space this large organized within a multi-level hierarchy, there is a vast space of possible event code aggregations.

\begin{table}[t]
  \centering
        \small
      \begin{tabular}{|l|r|r|r|r|}
      \cline{2-5}
         \multicolumn{1}{c|}{}
         & \multicolumn{1}{c|}{\bf Patients}
         & \multicolumn{1}{c|}{\bf Events} 
         & \multicolumn{1}{c|}{\bf Avg. Seq. Length} 
         & \multicolumn{1}{c|}{\bf Event Types} \\
      \hline
    \bf Minimum & 1,732 & 320,711 & 105 & 11,753 \\
    \bf Maximum & 8,360 & 1,136,681 & 185 & 15,376 \\
    \bf Average & 4,936 & 701,912 & 151 & 13,997 \\
      \hline
      \end{tabular}
    \vspace{0.1cm}
    \caption{Summary of event sequences returned by 12 \emph{Cadence} queries.}
    \vspace{-0.6cm}
    \label{tab:querystats}
\end{table}

\subsection{Design Requirements}

In prior work, the grouping of events by category as described above is typically performed as a pre-process \cite{du_coping_2017}.  As a result, analysts are typically forced to make educated guesses about how best to aggregate event types during the data pre-processing stage. If insights discovered during a visualization session cause an analyst to inquire about an alternative grouping, the analysis must be stopped, a new round of data pre-processing performed, and a new visualization session started on the newly processed dataset.  Because of the complexity of the underlying data, each iteration of this workflow can take a significant amount of time.

Through hands-on experiences collaborating with practicing health data analysts using a variety of event sequence visualization technologies, we have observed that there is a need for more flexible, dynamic approaches that enable users to explore alternative levels of event type aggregation as an integrated part of the visual analytics workflow, rather than as a pre-process. More specifically, we have identified the following key design requirements:

\begin{itemize}

\vspace{-0.15cm}
\item[R1.] Computationally determine a default context-appropriate level of hierarchical aggregation.
\vspace{-0.25cm}
\item[R2.] Provide users with high-level control over how aggressively the computational method groups events.
\vspace{-0.25cm}
\item[R3.] Provide users with the ability to explore the full type hierarchy regardless of grouping level, and to select alternative groupings.%
\vspace{-0.25cm}
\item[R4.] Interactive support for R2 and R3 within users' workflows.
\vspace{-0.05cm}

\end{itemize}

\section{Design and Algorithms}
\label{sec:methods}

The requirements outlined in the previous section motivate the development of new techniques for dynamic hierarchical aggregation.  This section begins with a brief overview of %
our visual analytics system for high-dimensional event sequence analysis, focusing on aspects that support the hierarchical aggregation process.  It then describes the key algorithms developed to enable these capabilities.

\subsection{Visualization Design}

The dynamic hierarchical aggregation techniques described in this paper have been developed within the context of \emph{Cadence}, a visual analytics platform designed for 
temporal event sequence analysis.  %

\subsubsection{Data Description and Defining Cohorts}

\emph{Cadence} enables users to define cohorts for analysis from large collections of longitudinal electronic health record data.  In these data sources, patients are represented with a combination of non-temporal attributes (e.g., gender and race) and temporal event sequences with hundreds or thousands of events per patient, capturing several years of medical history (e.g., diagnoses and procedures from specific dates).  Event type hierarchies provides multiple levels of abstraction over types of events (e.g., the ICD-10 coding hierarchy \cite{noauthor_icd_2018} for diagnosis codes).

The query interface, not shown, requires users to specify \emph{inclusion criteria} and \emph{outcome criteria}.  The inclusion criteria specify temporal event constraints for all patients to be returned by a query (e.g., the key diagnosis or procedure events, in order, that all patients returned by a query must have in their medical record).  In addition to determining which patients are returned by a query, the event constraints also define time windows of interest for each patient.  The outcome criteria specify temporal event constraints used to label each patient that meets the inclusion criteria with either a good or bad outcome (e.g., a bad outcome for patients who are eventually diagnosed with a particular disease).  
In this way, the cohort $P=\{P_i\}$ returned in response to a user query includes a set of patients, $P_i=\{\vec{a}_i,\vec{e}_i,v_i\}$ with attributes $\vec{a}_i$, a temporal event sequence $\vec{e}_i$, and an outcome $v_i$.  This representation is similar to outcome-labeled temporal event sequences used widely in prior work (e.g., \cite{gotz_decisionflow:_2014,wongsuphasawat_outflow:_2011,du_eventaction:_2016,zhao_visual_2018,mathisen_clear_2017}).

\begin{table}[t]
    \centering
        \small
        \begin{tabular}{|c|l|}
            \hline
            {\bf Symbol} & {\bf Definition} \\
            \hline
            $P = \{P_i\}$ & A cohort of $n$ patients\\
            $P_i = \{\vec{a}_i, \vec{e}_i, v_i\}$ & \makecell{A single patient with attributes $\vec{a}_i$, temporal event sequence \\ $\vec{e}_i$, and outcome label $v_i$}\\
            $\vec{v} = (v_1, v_2, \ldots v_n)$ & A vector of all patient outcomes \\
            $j$ & An event type \\
            $C_j=\{c_{j1},c_{j2},\ldots,c_{jk}\}$ & The $k$ children of event type $j$ in the event type hierarchy \\
            $\vec{t_j}=(t_{j1},t_{j2},\ldots,t_{jn}) $ & Binary length-$n$ event occurrence vector for event type $j$ \\
            $X^2_j$ & Informative metric for event type $j$ \\
            \hline
        \end{tabular}
        \vspace{0.03cm}
        \caption{A summary of the primary notation used throughout this paper.}
        \vspace{-0.7cm}
\label{tab:notation}
\end{table}

\subsubsection{Visual Interface}
\label{sec:interface}

A cohort is visualized using multiple coordinated views as shown in Figure~\ref{fig:teaser}.  First, the left sidebar in Figure~\ref{fig:teaser}(a) includes interactive bar charts and histograms used to summarize both categorical (e.g.,~gender) and continuous (e.g,~age) attribute variables, respectively. Right clicking on these charts enables users to revise the cohort's inclusion criteria by applying additional attribute constraints.  In addition, the left sidebar includes a sortable table of event types (e.g.,~both individual events such as diagnoses or procedures, and groups of these events from the type hierarchy) that occur at least once within the patient event sequences included in the cohort.  Users can sort the table by the number of sequences that include the event type, the total number of occurrences of an event type (an event type can occur multiple times for one patient), and the correlation between the occurrence of an event type and the outcome label.

Two additional linked visualizations are located on the right side of the interface.  This includes a Kaplan-Meier plot \cite{rich_practical_2010} in Figure~\ref{fig:teaser}(c), a traditional representation for ``survival'' data in the medical domain, which depicts the time of onset for the outcome variable.  Finally, the right sidebar also includes a scatter-plus-focus visualization showing event type prevalence and association with outcome.  The individual marks in the chart correspond to event types (both individual events, and groups of events from the type hierarchy) and are color-coded based on correlation with outcome.  In normal mode (Figure~\ref{fig:teaser}(d)), the circles are positioned by the percentage of patients with a matching event (the y axis) and the correlation of that event type or group with the outcome (the x axis).  

Critically, showing marks in the scatter plot for all possible event types and groups would suffer from severe over-plotting due to the vast number of choices, making it difficult for users to discover informative events or event groupings. %
To overcome this challenge, the chart first visualizes the density of all event types and hierarchical type groups (every node in the event type hierarchy) 
using a grayscale hexmap as a background (darker gray representing a higher number of event types).  %
It then employs a context-dependent importance measure to define a cut through the event type hierarchy which provides the most informative level of aggregation (\emph{R1}). Event types or groups along this cut are then visualized as marks within the scatter plot.  A slider located above the plot enables users to make global adjustments as to how aggressively the importance measure is used to group events (\emph{R2,R4}).  

Clicking on any of the event type marks transitions the visualization to a focused mode (Figure~\ref{fig:teaser}(e)) that enables users to navigate up and down the event type hierarchy for specific event type groups (\emph{R3,R4}). The focused view uses an optimization-based layout algorithm and scented glyphs to help users navigate the hierarchy.  The focus event type is shown using an enlarged circle, with its supertypes in the type hierarchy arranged above it along an x axis that encodes correlation of the event type with outcome.  Arcs connecting the event types enable users to visually trace the path to the root event type and see the changes in correlation at different levels of aggregation.  The x axis is repositioned to center around the focus event type and zoomed in to a narrower extent to support more detailed comparison.  Animated transitions emphasize the change in scale to users.  

Below the focused event type, a focused scatter plot shows marks for all child event types (direct descendants of the focused type in the type hierarchy).  The same focused x axis correlation scale is used to position the marks for both the supertypes and the child types.  As a result, users are able to make easy comparisons of correlations to outcome of event type groupings moving both up and down the hierarchy.  The y axis for the focused scatter plot corresponds to the proportion of event sequences containing an event type (as in the normal scatter plot).  However, the top of the axis is positioned just below the focused event type mark and the maximum value for the axis adjusted to be equal to the proportion of sequences containing the focused event type.  
This ensures that all child event types fit within the y axis range.

Any of the supertype or child type marks can be clicked to change the focus to the corresponding event type.  This enables users to precisely navigate the type hierarchy to explore alternative grouping levels (\emph{R3,R4}).  Because the focused mode only displays a small neighborhood of the full type hierarchy, visual scenting is used to help guide user exploration through the hierarchy. The scent is visualized using variable width triangles located beneath each circle.  Wider triangles suggest that the subtree located below the event type in the hierarchy exhibits high heterogeneity for correlation while a narrow triangle suggests that the subtree is more homogeneous. More details about the techniques used in the scatter-plus-focus visualization are provided in Section~\ref{sec:algorithms}. 

At the center of the interface is an interactive event sequence timeline which follows an interactive milestone-based design similar to DecisionFlow and subsequent systems \cite{gotz_decisionflow:_2014,law_maqui:_2018}.  In this view, \emph{milestone events} are represented with vertical bars whose height corresponds to the proportion of patient event sequences with a given event.  These vertical bars are linked with \emph{time edges} whose width encodes the average time between milestones for sequences that contain the corresponding event types in the required order.  As with milestones, the height of a time edge encodes the proportion of a cohort's patients whose sequences are represented by the edge.  

Both milestones and time edges are colored by the average outcome of the corresponding patient group using a red-yellow-green color scale.  This default scale uses red to represent poor outcomes and green to represent good outcomes, with the extent of the scale determined via the interactive legend above the timeline.  A red-yellow-green scale is used by default to align with the preferences and familiarity of the target medical user population.  However, alternative color scales are available to accommodate color-blind users.  

\begin{figure}[t]
  \centering
  \includegraphics[width=2.5in]{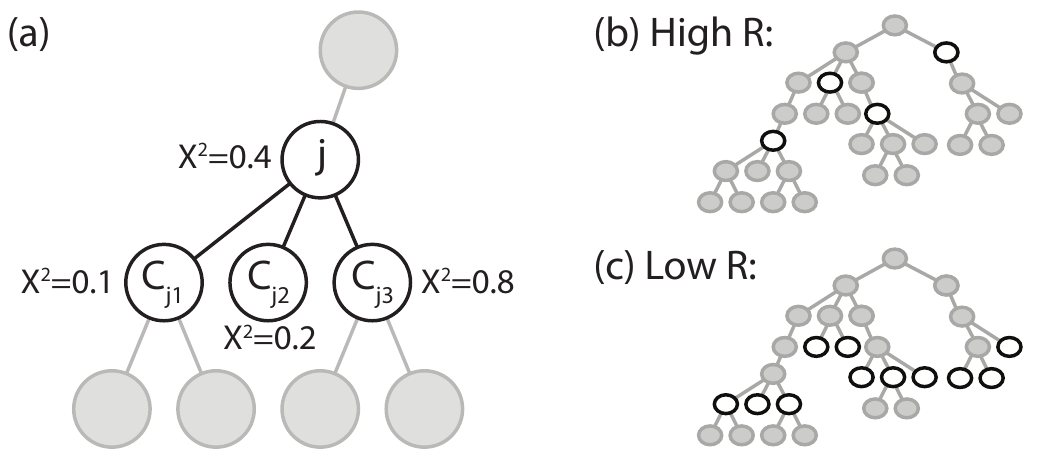}
  \vspace{-0.35cm}
    \caption{(a) The most informative cut through the event type hierarchy is determined by descending until the fraction of children more informative than their parent is below a threshold $R$.  (b) A higher $R$ results in a higher cut. (c) A lower $R$ will descend deeper into the hierarchy.}
  \vspace{-0.55cm}
\label{fig:r_example}
\end{figure}

In Figure~\ref{fig:teaser}(b), the timeline view displays a cohort of patients diagnosed with pain prior to being discharged from the hospital.  In addition, the visualization shows data for one year prior to the pain diagnosis.  The parallel paths at the start of the timeline show that slightly fewer than half of these patients were also discharged from a hospital visit in the year prior to the pain diagnosis, and that those patients had better outcomes than their ``not hospitalized before pain diagnosis'' peers.

Importantly, the timeline visualization enables the user to select either milestones or time edges to view more details about the patients and events that they represent. For milestones, data is displayed about events that occur at the same time as a selected milestone event.  For time edges, data is displayed for all events that occur in between the pair of milestones that define the edge.  Upon a new selection within the timeline, the charts in both sidebars are updated to show corresponding data.  Most critically, each time the selected section of the timeline changes, the scatter-plus-focus visualization is recomputed with an updated set of event groupings for the current context.  Animated transitions combined with event type search and highlighting enable users to compare event statistics across different timeline elements.  Finally, users can select an appropriate event type or type grouping from the scatter-plus-focus visualization to be added as a new milestone in the timeline.  Following the pattern of prior work \cite{gotz_decisionflow:_2014}, adding a new milestone enables exploratory analysis by causing the creation of new time edges for which statistics are recursively calculated and visualized.

\subsection{Key Algorithms and Interactions}
\label{sec:algorithms}

The interface described above leverages three key algorithmic solutions to support interactive control over the hierarchical event type grouping process.  This section defines the three algorithms and describes how they are used during visualization.

\subsubsection{Informativeness Measure for Hierarchical Aggregation}
\label{sec:informativeness_measure}

The scatter-plus-focus visualization provides users with information about both the prevalence of different event types, and the association between the patient outcome and occurrence of those event types. 
However, many event types have very low frequencies of occurrence which makes them less informative when looking for statistically meaningful patterns. An event type hierarchy can aggregate these low level events into higher level groups, resulting in fewer events with higher frequencies.  However, too much aggregation results in a loss of the information that analysts are seeking in their analysis (e.g., aggregating all the way to the generic root event of a hierarchy would eliminate all distinguishing information between events).   
In summary, visualizing events with either too little or too much aggregation can result in loss of information. We therefore define an informativeness measure which we evaluate for every node in the event type hierarchy each time the user's analytic context changes (i.e., via selections or the creation of new milestones in the timeline).
The results are used to determine a cut through the event type hierarchy representing the optimal (in terms of the measure) level of aggregation. The optimal groupings are then visualized (by default) within the scatter plot.

{\bf Measure Definition.} Given a specific cohort $P = \{P_i\}$ with events $\vec{e}_i$ and outcome $v_i$ for each patient $P_i$, an informative measure based on the chi-square statistic is computed for every event type in the hierarchy. We define a binary outcome vector for the cohort as follows:
\vspace{-0.2cm}
\begin{equation}
    \vec{v}=(v_1,v_2,\ldots,v_n)
\vspace{-0.1cm}
\end{equation}
where $v_i=1$ if the patient has the outcome or $v_i=0$ otherwise.

We then define a binary event type vector $\vec{t_j}$ for each event type $j$ in the event type hierarchy as follows:
\vspace{-0.2cm}
\begin{equation}
    \vec{t_j}=(t_{j1},t_{j2},\ldots,t_{jn})
\vspace{-0.1cm}
\end{equation}
where $t_{ji}=1$ if either type $j$ or a subtype of $j$ occurs at least once within $\vec{e}_i$, or $t_{ji}=0$ otherwise.  We emphasize that values in this vector are set equal to 1 if the type $t$ or any of its subtypes within the event type hierarchy are observed for a given patient.  Therefore, when $j$ is the root event type at the top of the hierarchy, the vector $\vec{t_j}$ will be a vector of all ones because all events are subtypes of the root event type.

\begin{figure*}[t]
  \centering
  \includegraphics[width=\linewidth]{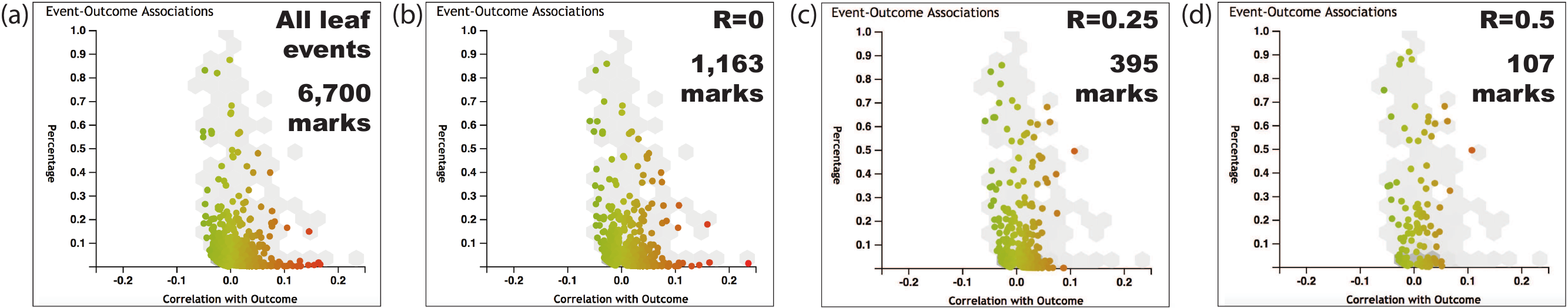}
  \vspace{-0.55cm}
    \caption{The level of aggregation is controlled by $R$, with higher $R$ values resulting in more aggressive event type grouping and fewer visual marks.}
  \vspace{-0.25cm}
\label{fig:simplication_scatterplots}
\end{figure*}

We then tabulate a contingency table based on the two previous binary vectors (each with length $N$) as follows:
\vspace{-0.1cm}
\begin{center}
    \small
    \begin{tabular}{|c|c|c|c|} \hline
        \diagbox[innerwidth=2cm]{Event}{Outcome} & 0 & 1 & Total \\ \hline 
        0 & $n_{00}$ & $n_{01}$ & $n_{0\cdot}$ \\ \hline 
        1 & $n_{10}$ & $n_{11}$ & $n_{1\cdot}$ \\ \hline 
        Total & $n_{\cdot 0}$ & $n_{\cdot 1}$ & $N$ \\ \hline 
    \end{tabular}
    \vspace{-0.1cm}
\end{center}
where for event type $j$, $n_{ab}=\sum_{i=1}^{N}(I[t_{ji}=a]I[v_i=b])$ and $I$ being the indicator function. We calculate a statistic based on the chi-square statistic for independence with a Yates correction for continuity:

\vspace{-0.1cm}
\begin{equation}
    X^2_j=\frac{N(max(|n_{00}n_{11}-n_{01}n_{10}|-N/2,0))^2}{n_{\cdot 0}n_{\cdot 1}n_{0\cdot}n_{1\cdot}}
\vspace{-0.1cm}
\end{equation}
where $X^2_j \sim \chi^{2}_{1}$\cite{Agresti:2002}, and $X^2_j=0$ if any term in the denominator is zero. This measure reflects the strength of the association between an event type and the outcome, with larger values representing  a stronger association.  Yates correction is used to prevent overestimation of rare event types (common in sparse data) due to assumptions behind the chi-square statistic \cite{yates_contingency_1934}. 
The chi-square statistic was selected as the basis of this measure because it is well-suited to quantify the strength of association between two binary variables sampled from the same population, and because it gives a p-value. However,  other methods, such as the Jaccard Index or other set-based statistics, could be used as the basis for alternative measures and could fit naturally within our overall framework.

{\bf Selecting Informative Event Type Groupings.}
Given a hierarchy of event types, and the importance measure $X^2_j$ which can be applied to any type $j$ within the hierarchy, we next define an algorithm that determines a cut through the type hierarchy such that the selected event types represent the most informative level of event groupings (\emph{R1}).

The algorithm for determining the most informative cut recursively traverses the event type hierarchy starting from the root event type.  At each event type visit, the algorithm compares event type $j$ with all of $j$'s children in the hierarchy. If $j$ is determined to be more informative than its combined children, then type $j$ is selected as part of the cut and its descendants in the hierarchy are no longer considered. If the child event types are considered more informative, the the process is recursively applied to each of the children one at a time to descend further into the hierarchy.  If the algorithm reaches a leaf node in the hierarchy, the leaf node is considered part of the most informative cut.

The key decision point in this algorithm is the comparison between the informativeness of type $j$ and the informativeness of its children $C_j=\{c_{j1}, c_{j2}, \ldots c_{jn}\}$.  Because $C_j$ can contain more than one event type, we define a one-to-many comparison criterion $R_j$ which is based on the proportion of children event types $C_j$ for which the informative measure exceeds the informativeness of $j$.  More formally:
\begin{equation}
\vspace{-0.2cm}
    R_j = \frac{\sum^{|C_j|}_{i=1}I[X^2_j<X^2_{c_{ji}}]}{|C_j|}
\vspace{-0.05cm}
\end{equation}

where $c_{ji}$ is the $i$th child of event type $j$. Type $j$ is classified as informative if either (1) it has no children or (2) $R_j \leq R$, where $0\leq R \leq 1$ can be specified by the user ($R=0$ being the default) via a slider in the user interface (\emph{R2,R4}). After determining the most informative cut, we then select events in the cut where $X^2_j>0$.
Intuitively, a lower value of $R$ generally selects events that are deeper into the event type hierarchy because it reflects a stricter criterion for stopping the hierarchy traversal algorithm. This results in less aggregation of event types, and therefore a larger number of lower level informative events.  In contrast, a higher $R$ will result in a cut of informative events that is higher in the event type hierarchy.  This produces a cut with fewer and higher level event types that result in a greater amount of aggregation.  This effect is illustrated in Figure~\ref{fig:r_example}.

The effect of adjusting the $R$ threshold with a realistic dataset containing an event hierarchy with 13,118 event types is shown in Figure~\ref{fig:simplication_scatterplots}.  The figure's first panel shows the result from choosing all leaf nodes in the type hierarchy (event types without children), which is the approach followed in prior work.  The other three panels depict the results when using R values of 0, 0.25, and 0.5  At the most aggressive level of aggregation, only 107 event types are displayed in the scatter plot.  This threshold can be adjusted interactively by users via the Hierarchy Simplification slider located above the scatter plot in Figure~\ref{fig:teaser}(d,e).

\begin{figure}[t]
  \centering
  \includegraphics[width=2.9in]{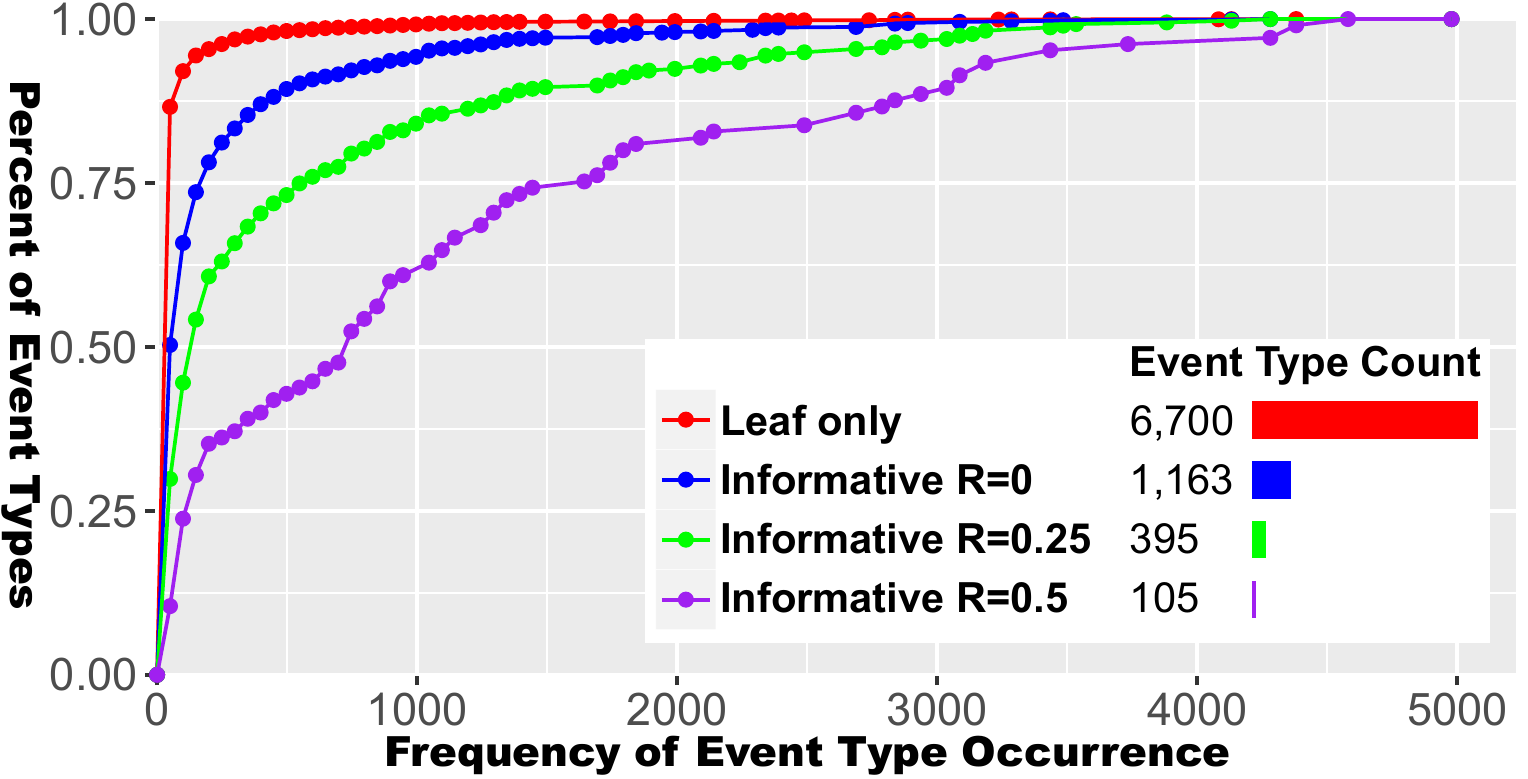}
  \vspace{-0.25cm}
    \caption{Using only leaf event types results in many low-frequency event types.  As this data from a test use case shows, hierarchical aggregation results in fewer and higher-frequency event types.}
  \vspace{-0.15cm}
\label{fig:freqplot}
\end{figure}

Figure~\ref{fig:freqplot} provides a deeper look at the effect of aggregation at these four different configurations (all leaves, R=0, 0.25, and 0.5).  These charts show that nearly all event types in Leaf Only mode have very few occurrences.  This makes the detection of meaningful patterns extremely difficult.  Increasing R values result in increasing amounts of aggregation that produce fewer event types with far greater frequency of occurrence.

We note that $X^2_j$ is influenced by the expected variance of event type occurrence frequencies in real world datasets.  Somewhat analogous to false negatives in traditional statistical measures, random fluctuations in these frequencies can potentially impact the selected event type grouping level. While computational techniques could potentially limit this effect (e.g., \cite{gotz_visual_2019}), the impact is mitigated by the fact that the cut provides only a starting point for aggregation, with users able to interactively explore alternative levels of grouping.

\subsubsection{Optimization-Based Layout for Focused View}

\begin{figure}[t]
  \centering
  \includegraphics[width=2.7in]{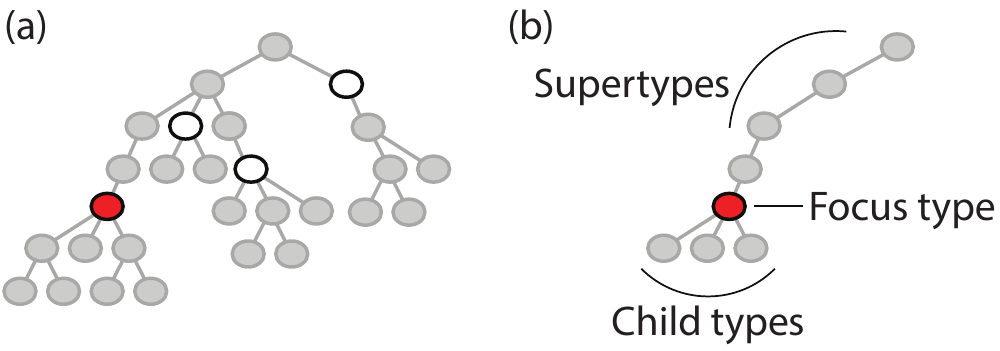}
    \vspace{-0.35cm}
    \caption{(a) Clicking on an event type in the scatter-plus-focus visualization transitions the chart to a focused mode. (b) The focused mode displays the selected event type, its children, and all supertypes up to the root of the hierarchy. All other types are hidden from view.}
    \vspace{-0.35cm}
\label{fig:focus_example}
\end{figure}

When users click on an event type mark in the scatter-plus-focus visualization, the chart transitions to a focused mode that enables users to explore up and down the type hierarchy,
as described in Section~\ref{sec:interface}. More specifically, the focused mode displays the focused event type, its supertypes up through the root of the type hierarchy, and all immediate children (see Figure~\ref{fig:focus_example}).

\begin{figure*}[t]
  \centering
  \includegraphics[width=\linewidth]{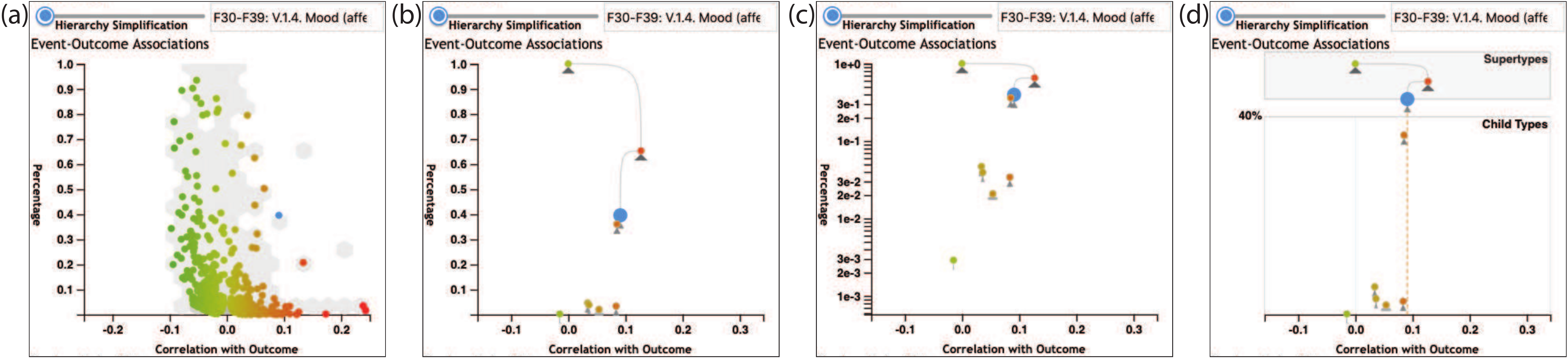}
  \vspace{-0.45cm}
    \caption{Clicking on an event type in (a) the scatter-plus-focus visualization transitions to a focused view.  (b) Maintaining the axes in the focused view results in overplotting, especially for the large number of low frequency events common to high-dimensional event sequences. (c) A log scale reduces overplotting for low frequency event types, but worsens problems for more frequent events and makes interpretation less intuitive. (d) An optimization-based layout maintains an intuitive scale while overcoming challenges of overplotting.}
  \vspace{-0.4cm}
\label{fig:optlayout}
\end{figure*}

Critically, navigating up and down the event type hierarchy requires users to see and directly interact with (via clicking) the individual event type circles in the focused chart. To overcome challenges of overplotting (see Figure~\ref{fig:optlayout}),  
a dual-view design was adopted.

This design positions the focused event type and its ancestors above an optimization-based scatter plot of the focused type's children.   As shown in Figure~\ref{fig:optlayout}(d), the top of the chart starts the path from the root event type, through all supertypes, to the focused event type. The y axis in this portion of the chart maps to the depth of the event type in the type hierarchy.  Below the focused type is a scatter plot of all children using an x axis that is centered on the correlation value for the focused type and aligned between both portions of the dual-view chart.  Vertical guide lines depict both zero correlation and the focused event type's correlation value to facilitate comparison between types.  

The y axis positions for the marks within the child type scatter plot are determined by an optimization-based layout algorithm that aims to balance two competing layout priorities: (1) marks should be positioned as close as possible to their ideal scatter plot location within the y axis scale, and (2) marks should not overlap.    For this reason, y axis positions closely approximate the actual percentage of event sequences that contain the corresponding event type, with adjustments made to avoid overplotting.

For a focus event type $j$, the layout process begins by assigning every child event type $C_{ji}$ to an initial position $(x_i,y_i)$ using the scatter plot's axis scales.  Then, vectors $\vec{x}$ and $\vec{y}$ are defined as the initial x and y positions for all marks, respectively.  The values in $\vec{x}$ are held constant and used to render the marks within the visualization.  However, the y positions are adjusted via a minimization process to obtain an optimized set of y positions $\vec{y}'$. The optimization minimizes the following cost function:
    \vspace{-0.2cm}
\begin{equation}
\begin{aligned}
    \argmin_{\vec{y'}} f(\vec{y_i}) & = \alpha\textcolor{Bittersweet}{\sum_{i=1}^{|\vec{y^\prime}|}\sum_{j=1}^{|\vec{y^\prime}|}\Omega(y'_i,x_i,y'_j,x_j)}
     + (1-\alpha)\textcolor{Plum}{\sum_{i=1}^{|\vec{y^\prime}|}|y^\prime_i-y_i|} \\
    where \quad & \textcolor{Black}{\Omega(x_1,y_1,x_2,y_2) = \max(0, d - \sqrt{(x_1-x_2)^2 + (y_1-y_2)^2})} \\
    s.t. \quad & \textcolor{Sepia}{y_{min} \leq y'_i \leq y_{max} \quad \forall y'_i \in \vec{y'}} \\
                            & \textcolor{PineGreen}{\sgn(y'_i - y'_j) = \sgn(y_i-y_j)} \quad \forall i,j \in [1,|\vec{y'}|]
\end{aligned} 
    \vspace{-0.1cm}
\end{equation}
where $d$ is the diameter of a mark in the plot, and  $\alpha$ is a tuning parameter used to balance between the competing elements of the cost function (overlap, and y-position distortion). 

Intuitively, the cost function includes two terms. First, \textcolor{Bittersweet}{an overlap term} sums the amount of spatial overlap between all pairs of marks.  A zero-cost layout would have no overlapping marks.  Second, \textcolor{Plum}{a distortion term} sums the amount of y-scale displacement applied to each mark. A zero cost layout would have no distortion to the vertical position of the marks.  The relative importance of these terms is controlled by a weighting factor, $\alpha$, which we set to 0.8 based on empirical observations that show it performs well in most cases.  

Two constraints are applied during the optimization process.  First, \textcolor{Sepia}{an extent constraint} requires that the optimized y axis positions ($\vec{y}'$) fall within the extent of the y-axis scale($y_{min}, y_{max})$.  This ensures that the optimized positions remain within the y-axis bounds of the chart.  Second, \textcolor{PineGreen}{an ordering constraint} ensures that the original ordering of marks along y axis is preserved in the final resulting layout (i.e., marks that occur more frequently are located above marks that occur less frequently).

The effect of the optimization can be seen when comparing the lower regions of Figure~\ref{fig:optlayout}(b,d).  The marks in the optimized version in (panel d) are more clearly separated out along the bottom of the chart.  This result increases legibility with less overplotting even when compared to the log scale version (panel c), while maintaining the approximate location of these marks along the bottom of the chart to correctly communicate that the corresponding event types rarely occur.

\begin{figure}[t]
  \centering
  \includegraphics[width=2.2in]{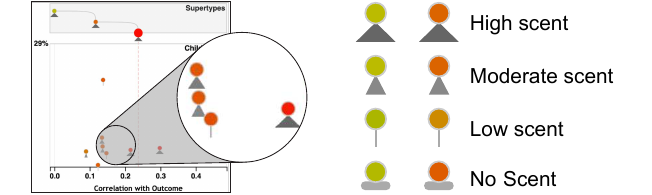}
  \vspace{-0.3cm}
    \caption{Scented glyphs help users efficiently navigate the type hierarchy.}
  \vspace{-0.35cm}
\label{fig:scenting}
\end{figure}

\subsubsection{Scenting}

The focused mode visualization includes marks for the focused event type, its supertypes, and its direct children.  Users can click on any super- or child type to change focus and navigate up or down the type hierarchy.  To help guide the user towards more interesting event types within the hierarchy during interaction, a scent value is calculated for all events and rendered as part of the glyph used to represent each event type. Intuitively, the scent is designed to highlight event types whose descendants in the type hierarchy have a wide range of correlations to outcome.  This property of an event type suggests that users might be interested in a lower level of aggregation that better separates event sequences by outcome.  Conversely, event types whose descendent types have more homogeneous associations with the outcome would be less valuable to disaggregate
The glyph for communicating the scent value is illustrated in Figure~\ref{fig:scenting}.  Event types that are at the bottom of the type hierarchy with no children cannot be expanded, and therefore are displayed with the ``no scent'' indicator instead of the normal triangular representation. 

The scent value for a given event type $j$ is computed from the correlation values for the entire subtree of event types below $j$ in the type hierarchy (not just the immediate children). The scent value is computed recursively, beginning with type $j$'s immediate children. The difference between the maximum and minimum correlation values for the event type's children is determined.  Then, the maximum scent for each of the children individually is determined.  The scent for $j$ is then equal to the maximum of either (1) the difference in correlation for $j$'s children, or (2) the maximum scent for $j$'s children. Leaf event types (with no children) are given a scent of zero by definition. As a result, the scent is equal to the maximum difference in correlation values for any set of peer event types within the subtree under $j$. 

This value is displayed as a glyph below each event in the outcomes view when focused on a specific event. As illustrated in Figure \ref{fig:optlayout}, the size of the glyph describes the magnitude of this scent value. There is a separate unique visual glyph if that event is an event without children. Although this scent value describes the heterogeneity in correlation values to provide the user with a general sense of what events have the most diverse children, it does not specify what level of aggregation the most diverse correlations are observed on. The user can click on an event to see which children of that event carry the diversity of correlation and make further inferences from there.

\section{Example Use Case and Domain Expert Interviews}

This section presents a use case that demonstrates how dynamic hierarchical aggregation can help support high-dimensional event sequence analysis tasks, and provides insights about the utility of these methods for health applications through domain expert interviews.

\begin{figure*}[t]
  \centering
  \includegraphics[width=7.00in]{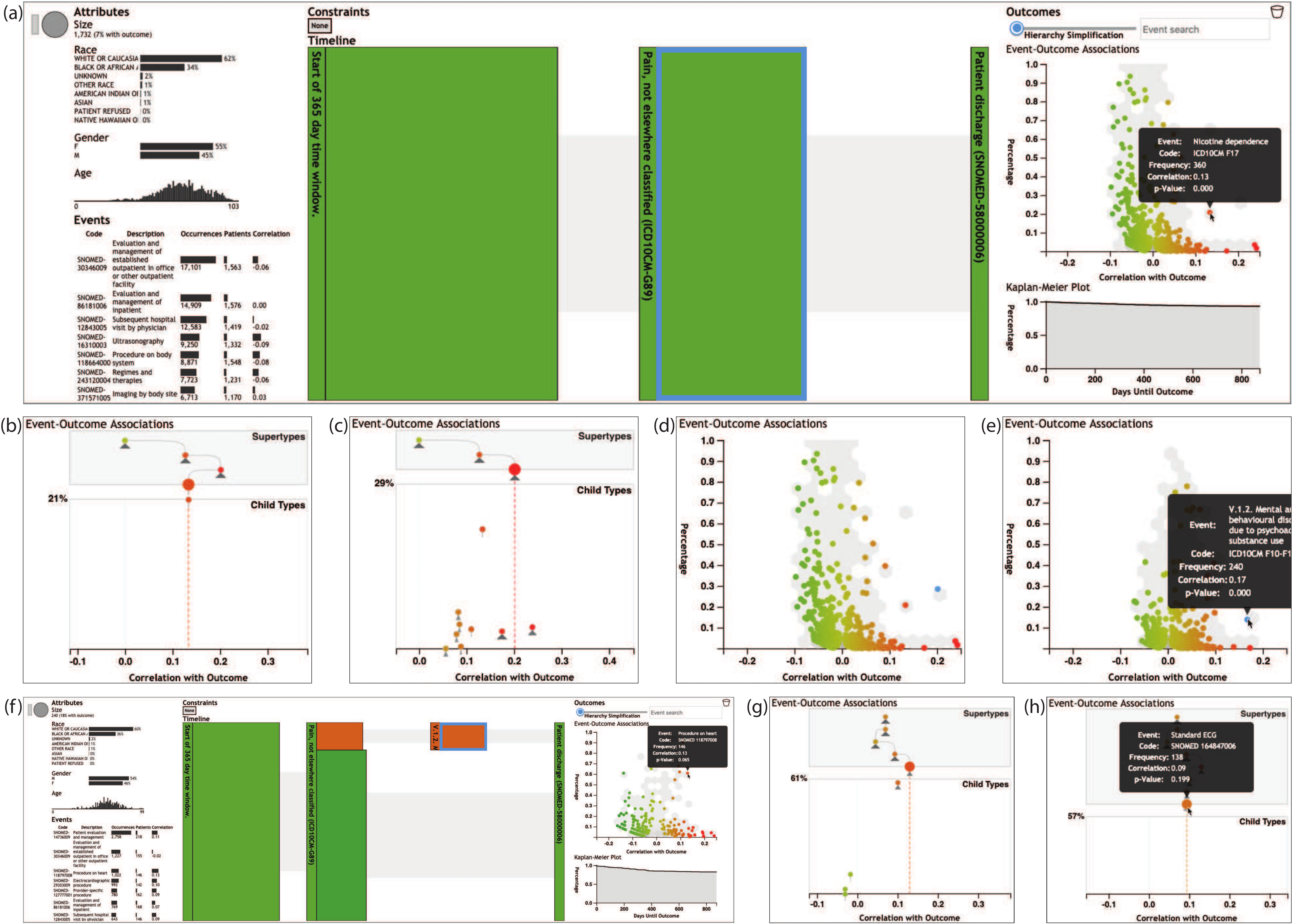}
  \vspace{-0.3cm}
    \caption{Screenshots from the use case described in Section~\ref{sec:usecase}.}
  \vspace{-0.35cm}
\label{fig:scenario}
\end{figure*}

\subsection{Example Use Case}
\label{sec:usecase}

Recognizing the value of analyzing and learning from medical practice, medical institutions and governments have invested in building large collections of of electronic medical data (e.g., \cite{murphy_serving_2010,fleurence_launching_2014,abernethy_rapid-learning_2010}) to support retrospective analyses that can inform policy making, quality of care, and medical understanding.  These ``real-world evidence'' efforts require analysts to sift through large and high-dimensional medical data to learn about which patterns or processes are associated with differences in medical endpoints, costs, or other outcomes of interest.  This has led to a growing interest in visual tools that can support interactive discovery over such data \cite{gotz_data-driven_2016}.

Consider the use case of an analyst working with data from our institution's own clinical data warehouse \cite{nc_tracs_north_2015} in an attempt to understand factors that are linked to opiate related disorders (i.e., a group of diagnoses related to addition or abuse).
Today, analysts working with this data would need to make a data requests from the data warehouse managers, who in response would return data files (e.g., SAS or CSV format) with extracts from the larger database.  A user would then need to manually pre-process these data extracts by cleaning and aggregating the tens-of-thousands of unique event types into a form that could be analyzed with prior visualization tools.  A hunch that alternative groupings would be more effective would require the analyst to leave the visualization environment, re-process the data files to generate the alternative groupings, and then re-visualize.  This process can be slow and cumbersome, interrupting the problem solving process with complex data manipulation tasks.

In contrast, Figure~\ref{fig:scenario} shows screenshots of how the system can be used for this type of analysis task.  The analysis begins with a query showing 1,732 patients who were discharged from a hospital after previously (during or before the hospitalization) having been diagnosed with pain.  The timeline in panel~(a) shows that the query also returned one years worth of events prior to the pain diagnosis.  Seven percent of the patients develop opiate related disorders after being discharged from the hospital.

Clicking the timeline segment representing time between the pain diagnosis and hospital discharge, the scatter-plus-focus chart is updated with the relevant statistics.  Nicotine dependence (a group of diagnosis codes with several subtypes) is a clear outlier.  It is quite prevalent (360 of 1,732 patients) and has a relatively strong correlation with opiate disorders ($\rho=0.13$). Clicking on the circle for this event type transitions the visualization to the focused mode shown in panel (b).  Here, users can see that there is a single child with very low scent.  However, moving up the hierarchy, the parent type has even stronger correlation and a large scent.  A tooltip reveals that this is a broader category of substance abuse (more than just nicotine), so the user clicks on the circle to revise the focus (panel~c).

The animated transition shows clearly that nicotine is this event type's child with the highest frequency (it is highest along the y axis), but not the most correlated with the outcome.  There are two less frequent subtypes of substance abuse that are relatively rare but exhibit higher correlation.  Due to the low frequency of those types, however, the user decides to stick with the higher level substance abuse event type grouping.  The user  locks the selection to this event type (highlighting it in blue) and clicks the chart background to return to the non-focused mode (panel~d).

Selecting the timeline segment prior to pain diagnosis, the scatter-plus-focus chart updates with new statistics and an updated set of most-informative event type groups (panel~e).  The user finds that the grouping remains associated with poor outcomes, but is less prominent (just 240 patients) and has weaker correlation.  The user therefore returns to the post-pain diagnosis timeline segment and adds the substance abuse event type group as a new milestone.  This results in an updated timeline (panel~f) which enables the user to continue searching.  In this case, the user finds heart procedures as a frequent and high-correlation event type group.  
Clicking on the corresponding circle, the focused chart shows that a more specific subtype has nearly the same frequency with high scent (panel~g).  The user can click down through several layers of the type hierarchy to discover that these are primarily (138 of the 148 heart procedures) ECG procedures (panel~h).  This event could also be added to the timeline, or the user could continue exploring the data.

\subsection{Domain Expert Interviews}
 
To gather qualitative feedback regarding dynamic hierarchical aggregation, we conducted hands-on demonstrations of the system and conducted semi-structured interviews with three medical experts.  The three participants were health-focused researchers with data analysis experience.  One participant was a medical doctor with joint clinical-research responsibilities, while the other two participants were PhD-level researchers.  Moreover, all three participants had prior experience with i2b2 \cite{murphy_serving_2010}, an NIH-funded web-based cohort selection environment that has been deployed to support data-driven health research at our university and at many other institutions around the world.  Experience with i2b2 was a requirement for recruitment to ensure that participants were within the target user population for the \emph{Cadence} system.
Each participant was interviewed independently, meeting with two study moderators for a one hour session. During the hour, the participants were given a brief introduction, followed by a demonstration of the system's key features. The participants were then interviewed by the study moderators in a semi-structured interview format.  During the interview, ad hoc data exploration was encouraged in response to participant curiosity.

\subsubsection{Thematic Analysis of Interview Findings}

The domain experts provided wide-ranging feedback during the interview sessions, addressing a variety of features. We present a thematic analysis of the interview findings, focusing on the themes that are most directly relevant to the methods presented in this paper.

{\bf Training required.} The users agreed unanimously that training is required to use the system.  One mentioned that it seemed complicated when they first saw the interface, but after being oriented ``it made sense.''  About the need for training, one expert emphasized ``I don't see a problem with that.'' Training is required for the existing i2b2 system, for example: ``i2b2 was complicated at first'' as well.  Speaking about both i2b2 and \emph{Cadence}, an expert remarked that these ``are for skilled users.'' Said another, the person using this is a ``superuser...willing to put in the time to learn the interface.'' One expert summarized this theme as follows: this is ``very cool, but there is a learning curve.''

The challenge of discoverability was mentioned by one user. ``People just have to know how to use [it].... They have to know they can manipulate the scroll bar to adjust the hierarchy.''  Similarly, the user asked ``how do I know if I can click'' on a circle in the scatter plot?  Training can help, but this is also an opportunity for future interface enhancements.

{\bf Benefits of automated selection of aggregation level.} The automated approach to suggesting the most informative level of aggregation ``was very useful.'' Another expert remarked that ``yes, [it was] very helpful,'' to have the system suggest a starting point.  One expert mentioned that a danger in an automated approach is that there was the potential that it would restrict the ability to explore the data. However, they didn't see that as a problem because ``you can control what level of aggregation is used'' referencing the slider which is mapped to the $R$ threshold described in Section~\ref{sec:informativeness_measure}.  They mentioned that the slider enabled them to ``control the simplification,'' and that it was very valuable and intuitive.

{\bf Intuitive navigation of the type hierarchy.} Users described the focused mode as ``intuitive'' and ``easy to interpret'' after brief training.  They found the hierarchy concept for event types natural, and intuitively understood what moving up or down that hierarchy meant in terms of the analysis.   ``The ability to go up in the hierarchy is really useful,'' mentioned one user.  After some exploration of the visualization on the screen at the time, another expert remarked ``clearly, different types of mental illness have different correlations with discharge. I found that very useful.''

With respect to the scenting, one expert found it ``very helpful.'' Another remarked that it was ``intuitive and easy to interpret'' after the brief training provided during the interview session.  ``I liked that... It was good.'' The experts found it easy to use the scenting to find where interesting variance was located within the type hierarchy.  There was a request, however, to display more detailed information about what led to a high scent score in response to a mouseover to avoid having to actually visit each highly scented event type.

{\bf Overall value of the approach.} A number of non-feature-specific comments were made more generally about the approach.  These were typically very positive as the prototype system provided many clear benefits to the domain experts. ``I really like your design.'' Said another user: ``This is very useful... for cohort studies.'' It enables you to ``instantly generate insights'' and is a powerful ``hypothesis generating application.'' One expert highlighted a discovery during the interview session:  when we ``saw the spike in the age distribution, [I asked] why?''  The system lets you look into it right away. 

This is a ``really powerful analytical tool'' said one expert.  ``This could be a powerful tool'' said another, ``it has all the function that people could imagine....  I didn't know that a user interface could do this much. Could go this deep. [Could help you] choose event levels.''

\vspace{-0.1cm}
\section{Conclusion}
\vspace{-0.1cm}
This paper presented a new visual analytics approach for dynamic hierarchical dimension aggregation during high-dimensional temporal event sequence analysis.  It overcomes limitations in prior work by enabling users to interactively and intuitively explore and define group type aggregations as part of the analysis workflow rather than as a pre-process.
The approach leverages a pre-defined event type hierarchy to computationally quantify the informativeness of alternative levels of grouping given the current analytical context.  This information is then visualized to give users the ability to interactively explore alternative groupings and select the most appropriate level of grouping to use at any individual step within an analysis.  This is made possible via (1) a measure of informativeness that can be applied to individual event type groupings; (2) an efficient and tunable algorithm for determining the most informative levels of aggregation from within a large type hierarchy; and (3) a novel scatter-plus-focus visualization with scenting and optimization-based layout to help users explore the type hierarchy to compare alternative levels of aggregation. Although these contributions have been implemented and evaluated (through domain expert interviews) as part of a medical data analysis tool, there is potential applicability to a broader range of similar problems.

While the results are promising, there remain several areas for future work.  For example, visualization methods that afford more flexible groupings beyond those strictly defined within the type hierarchy would very useful within the medical context.  Leveraging ontologies rather than tree-based hierarchies could help support this type of flexibility.  Another possible topic for future work is making it easier to re-use groupings from one part of an analysis in later stages.  Consistent grouping is often important, and better interactive support for this would be valuable.  Interface improvements to support discoverability of advanced features would also be useful.

\acknowledgments{
The research reported in this article was supported in part by a grant from the National Science Foundation (\#1704018).}

\bibliographystyle{abbrv-doi}

\bibliography{websites.bib,references.bib,sections/references.bib,zhang.bib}
\end{document}